Wikipedia: an opportunity to rethink the links between sources' credibility, trust and authority

Gilles Sahut[1] and André Tricot[2]

School of Education, University of Toulouse, Toulouse, France

56 avenue de l'URSS

31 078 Toulouse cedex - France

1 Laboratoire d'Études et de Recherches Appliquées en Sciences Sociales (LERASS)

Direct comments to: gilles.sahut@univ-tlse2.fr

2 Work and Cognition Lab., CLLE CNRS

Andre.Tricot@univ-tlse2.fr

**Wikipedia: an opportunity to rethink the links between sources' credibility, trust and authority**

Abstract

The web and its main tools (Google, Wikipedia, Facebook, Twitter) deeply raise and renew fundamental questions, that everyone asks almost every day: Is this information or content true? Can I trust this author or source? These questions are not new, they have been the same with books, newspapers, broadcasting and television, and, more fundamentally, in every human interpersonal communication. This paper is focused on two scientific problems on this issue. The first one is theoretical: to address this issue, many concepts have been used in library and information sciences, communication and psychology. The links between these concepts are not clear: sometimes two concepts are considered as synonymous, sometimes as very different. The second one is historical: sources like Wikipedia deeply challenge the epistemic evaluation of information sources, compared to previous modes of information production. This article proposes an integrated and simple model considering the relation between a user, a document and an author as human communication. It reduces the problem to three concepts: credibility as a characteristic granted to information depending on its truth-value; trust as the ability to produce credible information; authority when the power to influence of an author is accepted, i.e. when readers accept that the source can modify their opinion, knowledge, and decisions. The model describes also two kinds of relationships between the three concepts: an upward link and a downward link. The model is confronted with the findings of empirical research on Wikipedia in particular.

**Introduction**

"Those opinions are 'generally accepted' [*endoxa*] which are accepted by everyone or by the majority or by the philosophers-i.e. by all, or by the majority, or by the most notable and illustrious of them" noted Aristotle 24 centuries ago. The alternatives mentioned by the Greek philosopher are today more relevant than ever due to the recent evolution of the document. The freedom and ease of publishing on the internet come at a cost. Information seekers are faced with a wealth of sources from multiple social actors. It results in uncertainty about how credible information is, when it's not always published by gatekeepers traditionally tasked with producing and filtering information for the public, while respecting set processes and norms. Moreover, in the context of online publishing (blogs, wikis, social networks…), online document have new formal features. As a result, the credibility criteria usually used in the world of print media changed (Sundar, 2008; Jessen, Jørgensen, 2012).

This paper is focused on two scientific problems on this issue. The first one is theoretical: the semantic ambiguity of credibility, trust and authority in Library and Information Sciences, resulting in a lack of accuracy when considering the epistemic evaluation of information sources. The second one is historical: the social web deeply challenges the epistemic evaluation of information sources, compared to previous modes of information production. Our goal is to discuss these three concepts and their relationships, to propose a model of the epistemic evaluation of information sources. We use the case of Wikipedia to illustrate how credibility, trust and authority can be challenged today, but also how Wikipedia changed its way to produce information in order to increase its credibility, trust and authority.

**Credibility, trust, authority: three key concepts at the core of the epistemic evaluation of information sources**

Since Aristotle, credibility, trust and authority are three key concepts widely discussed in philosophy, epistemology and, more specifically, in philosophy and sociology of science. Credibility, trust and authority are involved in almost all our relationships with other human beings, with discourses and with knowledge! We won't discuss these issues here. Our paper is focused on these concepts as they are used in empirical and theoretical literature in library and information sciences. Several studies focused on the way individuals evaluate online information and its credibility. Other studies – probably less – were conducted regarding the strategies used by online sources to earn their readership's trust. In those studies, the terms "credibility", "trust", and "authority" were frequently used. Those terms prove challenging to define. It is established that the three concepts of authority, credibility and trust are all multi-dimensional, complex, and are conceptualized differently (Kelton, Fleischmann, Wallace, 2008; Rieh, 2010)[1].

*The credibility of information*

Empirical studies devoted to the concept of credibility arose in the context of pioneering research led by Carl Hovland and his colleagues at Yale, which started during the Second World War and became the core reference of numerous studies in communication (Metzger et al. 2003). A lot of research has been published in this field, mainly about the credibility granted to media, sources and messages, but also about their effects on behavior and opinions.

In the library and information sciences (LIS), the seeds of research on credibility were planted by studies focusing on information behavior and more specifically on relevance judgements made by the users of an information-seeking system. During the 1990s and 2000s, studies proved that in order to choose documents and information while researching, individuals did not use topicality only, but a great variety of other criteria as well, for instance credibility and reliability (Barry, Schamber, 1998; Maglaughlin, Sonnenwald, 2002). The enormous growth of Web use led to the proliferation of studies on credibility assessment because individuals are faced with an information environment with fewer editorial control mechanisms than in the traditional print environment (Rieh, 2010). This subject brings together the fields of LIS, media studies, social and cognitive psychology and human-computer interactions, even though the theoretical and methodological options prove dissimilar. Despite the diversity of studies on the topic, or because of it, credibility remains hard to define. In Rieh's words: "credibility has been defined along with dozens of other related concepts such as believability, trustworthiness, fairness, accuracy, trustfulness, factuality, completeness, precision, freedom from bias, objectivity, depth, and informativeness" (Rieh , 2010: 1337). To cope with this abundance of terminology, it seems best to return to a simpler definition. Indeed, Oxford Dictionaries define credibility as "the quality of being trusted and believed in". Furthermore, since Hovland and al. (1953), research has converged on two attributes categories used to assess the credibility of a source or of information (Fogg et al., 2003; Rieh, 2010, Choi, Stvilia, 2015). On the one hand, expertise, linked to the acknowledged skills of the source such as its status, qualification, or experience; on the other hand, its trustworthiness. It is linked to the moral qualities seen in the source, whether it is well-intentioned, truthful, and unbiased.

---

[1] It is obvious that literature about credibility, trust and authority in philosophy of science and in library and information sciences are related. For example, what we call the "downward model" (where authority ensures trust, that ensure credibility) has been challenged first in philosophy of science by Merton for example (Sztompka, 2007) and then in library and information sciences, in very close ways. We won't make comparisons between these disciplines here, because of a lack of place, but we are going to use different works from different academic disciplines to discuss our three concepts and their relationships.

*Epistemic trust*

We will discuss only one type of trust, that is relevant when seeking and using information: epistemic trust. We are concerned with epistemic trust when we have to believe others to get information on the world (Origgi, 2004). It is characteristic of knowledge transfer situations – including information seeking – which involve a knowledge discrepancy between the knowledge-seeking subject and the potential source of knowledge.

Trust is involved in situations where individuals feel uncertainty and vulnerability (Kelton, Fleischmann, Wallace, 2008). These two feelings are characteristic of the information-seeking process. Indeed, this process is triggered by the realization of a need for information, which in turn comes from the desire to reduce uncertainty regarding a problem in a misunderstood situation. In Kulthau (1991)'s Information Search Process model, uncertainty is linked to the first stages of research (initiation, exploration). Vulnerability springs from the potential damage that can be caused by using inaccurate information, leading to inappropriate decisions. On the web, the feelings of uncertainty and vulnerability are increased due to the absence of global norms assuring the quality of the available information. In light of this, the trust that is granted to information sources is a factor that diminishes both feelings.

When the ways of determining trust in a source of information are taken into consideration, the closeness of the link with credibility as a concept appears more clearly. A cognitive approach to trust assumes that the subject undertakes a rational assessment of the epistemic stance (Origgi, 2008). It is then possible to determine its capacity to "tell the truth" (its expertise) and its intentions to do so (intellectual honesty, loyalty, respect of their duties and responsibilities). Thus there are two dimensions that determine the credibility of a source: an epistemic norm, and a moral one.

*The authority of a source*

Patrick Wilson defined in 1983 the notion of cognitive authority as an individual accepting and acknowledging as appropriate the intellectual authority of a source (human, media, document) in order to get new knowledge (Wilson, 1983). Not only are these sources deemed credible, they are granted a higher level of expertise. They are indeed recognized as able to offer a state of the matter regarding a topic, to assess the qualities and flaws of a thesis, to give more or less value to existing opinions and to indicate what should be believed. In other words, a cognitive authority has a form of reflexivity in a given field of knowledge, which explains that it is held as a favored source.

Wilson's theory should be read in light of Bourdieu (1977) who proposed an analysis of the social mechanisms involved in granting authority to speech. According to the French sociologist, it comes from the author, who is considered a "spokesperson" in the sense that his authority was delegated to him by an institution. As an illustration to this delegation, Bourdieu uses the image of the *skeptron*, a scepter imperatively required to speak in public in ancient Greece. This thesis shows the benefits of emphasizing the power given by institutions to the social actors who represent them. It is associated with the idea that there is a hierarchy of sources and speech depending on their legitimacy, which itself comes from being part of an institution. In that regard, Bourdieu's analysis is not all that far from that of Wilson, who also showed the influence of institutions in how cognitive authority is built.

The concepts of credibility, epistemic trust and cognitive authority are used in different models about the truth value of information.

**Previous models**

Several models were developed to describe how credibility assessments are made on the web. In the Prominence-interpretation theory (Tseng, Fogg, 1999 ; Fogg, 2003 ; Fogg et al. 2003), credibility assessments occur when users notice an element of a website (prominence) and use it to infer the credibility of this website (interpretation). It turns out that the criterion that influences such assessments most frequently is the design look of the website (layout typography, white space, images, color schemes …), followed by structure of information (how information is organized, how easy it is to navigate), the information focus and then, the underlying motive of the site or the institution sponsoring the site. Burkell and Wathen (2002) used Fogg's conclusions as well as other studies on the same subject to design a model of the credibility of a website and of the decisions associated with it. In this model, the user makes an assessment in several consecutive steps. She or he makes a first assessment based on the superficial characteristics of the website: appearance, interface design and Web site organization. Then the user assesses the source (expertise, competence, trustworthiness, credentials….) and the message accuracy, currency, and relevance to her or his need. If this assessment is favorable, she or he finally makes a more in depth analysis of the content of the website. If in those steps the negative assessments prevail, the user leaves the website to seek another source of information. For Jessen and Jørgensen (2012), the need to produce a new model is justified by the continued evolution of online environment. The user of the social web (social networks, blogs, wikis,…) does not necessarily have the usual criteria regarding the expertise and trustworthiness of the source to determine the credibility of information. However, she or he can use a sort of social validation in the form of the assessment of other users (Likes on Facebook, followers and shares on Twitter, etc.) as well as the more traditional sign of being part of an institution.

Other authors have proposed models focused on the concept of epistemic trust. Kelton and al. (2008)'s model emphasizes the central role of trust, which is considered as a key mediating variable between information quality and information usage. It is granted depending on characteristics identified by the user: the skills of the source, its positive intentions, its ethical qualities and its predictability, i.e. its capacity to meet the expectations created by the source's previous behavior and the social role expected of it. Trust is also built up by other factors such as the context in which the trust is embedded. The propensity to trust as a purely psychological attribute, and the level of social trust in the recipient. The 3S model (semantics, surface, and source features of information) rests on empirical studies (Lucassen, Schraagen, 2011; Lucassen, Schraagen 2012). Depending on circumstances, the information seeker can either use his domain expertise to assess the semantic contents of a document, or their information skills to take signs of credibility into consideration, or their past experience with the source of the information. Additionally, it was observed that trust in information is influenced by trust in its source, trust in a source is influenced by trust in a medium, trust in a medium is influenced by a propensity to trust.

The concept of authority is also used to develop models and empirical studies on the truth value of information. Rieh used Wilson's concept of cognitive authority to propose a model of the selection of information on the web (Rieh, 2002). Much like Burkell and Wathen's model (2002), it notes that assessing an online document in a process in several steps. But unlike it, it indicates that that assessment process starts before the web page is read. This type of judgement is called "predictive" and it shows the expectations of the user. In other words, reader builds a representation of the source before even seeing it. The evaluative judgments are, in turn, made when the webpage is being read. If the two judgements match, then the document will be used. Otherwise, the user will stop reading. Previous knowledge of online sources influences the nature of the predictive judgment. The information-seekers rely first and foremost on their own experience with documents to trust sources of information. But they also use recommendations or opinions from colleagues, friends, as well as magazines, articles or even television commercials. The concept of authority has also been a framework for empirical studies. It was used to identify influential sources for social groups, and the foundations used to build this authority (Rieh, 2002; Mac Kenzie, 2003; Savolainen, 2007) or to question it. For instance, Neal and McKenzie studied how bloggers debated and refuted the authority of medical sources regarding health issues (Neal, McKenzie, 2011).

**A theoretical and historical problem**

Concepts of credibility, epistemic trust and authority of sources give rise to multiple research and models in library and information sciences. However, we are confronted with a theoretical problem. The ambiguities of these concepts do not help to have a unified approach to this area of knowledge. We develop this point and then address the historical problem: how to understand the questions of credibility, trust and authority posed by new sources such as Wikipedia.

*Ambiguity and confusion between concepts*

We note that credibility, trust and authority all refer to a particular dimension of approaching a piece of information: Believing in its truth-value. Thus the three concepts refer to what could be called epistemic judgments. They seem to overlap somewhat while still remaining distinct and as a consequence are not interchangeable. The semantic relationships between them seem less than clearly established. In some studies, these different notions orbiting around credibility are used almost as synonyms. For instance, Watson choses to equate credibility, reliability, accuracy, authority, quality of information and trust in information (Watson, 2014). Tsen and Fogg have noted the semantic problems with credibility and chose to use of credibility as a synonym of believability while trust is used as that of dependability (Tseng, Fogg, 1999). In other studies, credibility is sometimes a mean to operationalize trust (e.g. Menchen-Trevino, Hargittai, 2011). In others, it is trust that is a means to operationalize cognitive authority (Rieh, 2002).

A similar problem can be found in the theoretical models we mentioned. Some models of credibility link this notion with authority (Burkell, Wathen 2002; Jensen, Jørgensen, 2012) but not explicitly with trust. In the three S model, the concept of trust incorporates that of credibility (Lucassen, Schraagen, 2011). For Kelton et al. (2008), credibility is too narrow of a representation of trust. Using the concept of trust is to be favored as it allows one to take advantage of the many theoretical advances on the subject of this concept. We note that in the two models by Lucassen and Schraggen and Kelton and al. which focus on trust, the authority of the sources is only seldom taken into account. The same can be said of the Prominence-Interpretation theory which is focused on credibility (Fogg, 2003).

Our literature review confirms the semantic ambiguity of LIS about credibility, trust, authority. Empirical and theoretical literature in LIS does not offer definitions or clear models linking these three concepts that seems essential to understand the truth-value that individual give to information. Nevertheless it seems that the authority of a source derives from its attachment to knowledge institutions. Its power of influence is considered as legitimate because it comes from or is associated with a knowledge institution. We use this term to refer to legitimate social groups and institutions that have the mission to produce knowledge and/or to disseminate it. This general term includes scientific research, schools, libraries and museums, each of course with their specificities and complex relationships with one another. This kind of authority generates trust in a source. At first glance, however, Wikipedia does not appear to fit into this framework.

*The case of Wikipedia*

Traditionally, the most prestigious encyclopedias are from reputed editors and authors, who guarantee their authority (Rasoamampianina, 2012). They are also associated with academies, scholarly societies and universities to take advantage of their reputation (Sundin, Haider, 2013, Yeo, 2001). When it started in 2001, Wikipedia had neither the support of institutions of knowledge nor the contributions of known experts. It relies on volunteers and a process of open collaboration (Forte, Lampe, 2013). Thus this project does not comply with the usual norms for creating an encyclopedia. Yet it became one of the ten websites with the most traffic in the world and the most read online encyclopedia.

Studies emphasize that Wikipedia became part of that informational practices of an important part of the population, so much so in fact that the number of times that the articles were read or modified enable one to reliably predict the propagation of flu or dengue epidemics several days in advance (Generous et al. 2014) or the success of a movie in theaters even before its release (Mestyán, Yasseri, Kertész, 2013). Young people are shown to use it frequently (Flanagin, Metzger, 2010; Selwyn, Gorard, 2016), often with the search engine Google (Head, Eisenberg, 2010; Colón-Aguirre, Fleming-May, 2012). The majority of Wikipedia's users judge their experience as positive (Lim, 2009; Head, Eisenberg, 2010), including in terms of accuracy of the information (Sahut, 2014a). This factor contributes to the upward authority process of the source. However, despite this frequent usage and experiences in majority judges as satisfactory, trust in Wikipedia remains average (Julien, Barker, 2009; Watson, 2014; Georgas, 2014). How can we understand what seems a paradox?

Studies were conducted to understand the degree of trust in Wikipedia and other encyclopedias such as Britannica (Kubiszewski, Noordewier, Costanza, 2011; Flanagin, Metzger, 2011). The articles shown as belonging to Wikipedia were deemed less credible than those from Britannica. These experiments show that it is not due to the fundamental quality of the articles. The users were influenced by the reputation of the source. Britannica benefits from the credit it accumulated over its long history and the recognition from institutions of knowledge while Wikipedia's development model is deemed less reassuring. Let us remember that Wikipedia was the object of much criticism from academics (Reagle, 2010). Some of its critics saw it as a sign of mediocrity triumphing over expertise (*e.g*. Keen, 2007). Wikipedia's reputation turned out mostly negative among teachers and many of them advised their students against using it for academic projects (Francke, Sundin, 2012; Knight, Pryke, 2012; Purcell et al., 2013). Many studies show that young people take into account the negative opinion of their teachers on Wikipedia, which can lead them to abstain from quoting them as a source in their school projects (Lim, 2009; Head, Eisenberg, 2010; Watson, 2014).

Therefore Wikipedia has to demonstrate its credibility. For Jessen and Jørgensen (2012), credibility of information on Wikipedia comes from a process of collective validation, in that if a statement remains, then it was accepted by the other contributors. This is credibility from multiplicity a specific way of seeing credibility which asserts that collaborative efforts from multiple authors must lead to accurate information (Francke, Sundin, Limberg, 2011). This belief is deeply embedded in Wikipedia's community, but it's not the only one. In the English (Forte, Bruckman, 2005) and Swedish (Sundin, 2011) Wikipedia, the importance of the "Verifiability" and "Citing sources" rules was noted. Linking a statement to a source is a way to certify its credibility. In a way, referencing became the *skeptron* that legitimizes inserting an encyclopedic statement. This referencing effort is an important part of the contributors' activity. Analyzing the debates among the French Wikipedia community allows for a better understanding of why referencing rules were established (Sahut, 2014b). They were adopted between 2005 and 2007 when much criticism of Wikipedia was brought forth. The Seigenthaler case showed how fragile the project was (Reagle, 2010). When he read his page on the English-speaking Wikipedia, this American journalist saw he was allegedly involved in the assassination of the Kennedy brothers. He denied vigorously, threatened to sue, and denounced Wikipedia's editorial system. Historians, philosophers and linguists tried warning the public through the media of the epistemic dangers of this encyclopedia. The community realized the limits of a credibility solely relying on internal collective validation. To them, referencing is a process that compensates for the uncertainties inherent to open-editing, to editor anonymity, and the lack of a centralized and expert body with the mission to validate articles (Sahut, 2014b).

**A model linking authority, trust, and credibility (ACT model)**

As we have seen previously, the relationships between concepts of credibility, epistemic trust and cognitive authority have not been clearly established in the scientific literature. Furthermore, nowadays the authority of a source appears not only to come only from the institutions of established knowledge. Wikipedia is emblematic of this phenomenon. Thus, we propose a new model which links credibility, epistemic trust and authority of a source.

*Definition and semantic links between credibility, trust and authority of a source*

In continuity with the studies mentioned above, we propose definitions for the concepts of credibility, epistemic trust, and authority of a source.

- Studies on the credibility of a source can be found frequently in scientific literature. However, the judgments of this type are ultimately on the information itself, and the analysis of the source can only reinforce this judgment. We propose to define credibility as a characteristic granted to *information* depending on its truth-value.
- Trust characterizes a relationship in which a recipient (or a reader) recognizes that a *source* is able to produce credible information.
- The authority of a *source* corresponds to acknowledging and accepting its power to influence, that is to say the reader accepts that the source can modify his or her opinion, knowledge, and decisions. It guarantees trust, and indicates that a source is favored in a field of knowledge, when there are several available sources. The authority of a source is guaranteed by social mechanisms.

In order to supplement these definitions, we will attempt to specify the relationships between these different concepts from the point of view of the user.

- The credibility of information is often inferred from trust in a source.
- A source that regularly produces credible information is likely to favor the creation of a relation of trust with the user.
- A source that acquires the trust of an important number of people gets more authority.

We will set apart two possible origins of the authority of a source: the downward process of recognizing authority and the upward process of building up authority.

At the inception of the downward process is a source with institutional authority. But the role of all these institutions is to guarantee the epistemic value of information – *i.e.* the belief that it is true – that it produces through sources, whether human or in the form of documents. As Bourdieu (1977) suggests with the image of the *skeptron*, the social status of the source is recognizable from features of the document (titles of the author, the institution it belongs to, the nature of the editor, etc.). When the users recognize this authority, it supports the creation of a relation of trust and reinforces the credibility of information. Thereby is created a heuristic of judgments of credibility (Metzger, Flanagin, Medders, 2010).

Building authority can also be an upward process. Indeed, we estimated that the repeated experience of the credibility of a source favors the creation of a relation of trust with its users. If such users become numerous, it can result in social recognition. The upward authority process can thereby take place on a more or less expansive social scale (for instance on the scale of a scientific community, a professional community, or fans).

*Authority, trust and credibility (ATC) model*

We propose to integrate these concepts and the links between them in a communication model. We estimate that they can all be used to describe a dimension of a situation of communication between on the one hand an information producer, that we will call author or designer, and on the other the receiver, seeker, selector and processor of information (in short, a reader or user).

We propose to adapt Sperber and Wilson's relevance theory to documentary communication (Sperber, Wilson, 1986). According to Wilson and Sperber, most human communication, both verbal and non-verbal, is the expression and recognition of intentions. Dialogue works on two levels: the informative intention, to inform the recipients of something; the communicative intent, to inform the recipients of this informative intention. "According to this theory, […] a communicator provides evidence of her intention to convey a certain meaning, which is inferred by the audience on the basis of the evidence provided" (p. 249). The recipient fills the information gap produced by the communicator because he/she recognizes that the communicator tries to be relevant. The filling of the gaps dialogue is based on the knowledge that each one has of the other, on the knowledge that each one knows that of the other knows, etc. When designing a document, the author actualizes an intention to inform by targeting a readership. She or he also has an intention to communicate, to ostensibly inform his readership of his intention to inform. In this theory, the target (audience, reader…) determined the intentions and representations of the sender / author by inferring.

By focusing solely on the truthful quality that the author means to grant the information, we would say she ostensibly shows her intention to "tell the truth". In our model, it can be actualized on two levels (fig. 1):

- by giving credibility to information; for instance, by stating information that the author thinks believable for the targeted readership;
- by displaying signs in the source that allow to establish or enhance a relation of trust with the targeted readership. The author can thus show signs showing trustworthiness and expertise. For example, he or she can remind the reader of his/her qualifications, profession, publications, experience in the relevant domain of knowledge, etc. In the context of documentary communication, he or she benefits through the halo effect of the trust accumulated by social actors and institutions involved in realizing and validating his / her document and in the case of scientific writings, of being quoted or cited by other researchers. These clues embed in the document contribute to what we called the downward process of authority. Being associated with socially recognized institutions of knowledge in a factor of legitimization of the source, ensures trust in the source and the credibility of the information.

Let's now switch to the perspective of the reader. We will abstain from participating in the various debates held in the context of theories of communication in the 20th century over the processes of reading/interpreting/receiving messages. We will merely postulate that in the context of seeking information, the reader (or user) has a margin of interpretation with the document she or he selects, reads, and transforms into a source. Referring to Wilson and Sperber's theory, we will say that the reader can assess the credibility of information by identifying the intentions and expertise of the sender / author (fig. 1). To this end, their evaluative activity can be on two levels:

- The evaluation can be on the level of the information itself. The reader can evaluate the information using their own knowledge to assess its likelihood or evaluate whether it matches with their pre-existing beliefs and opinions;
- The evaluation can be on the level of the source. In order to trust the source, the reader can rely on recognizing signs that symbolize social qualifications and indicate expertise and/or trustworthiness.

Through these processes, the reader can infer whether the author intends to "tell the truth" and whether he is capable of doing so.

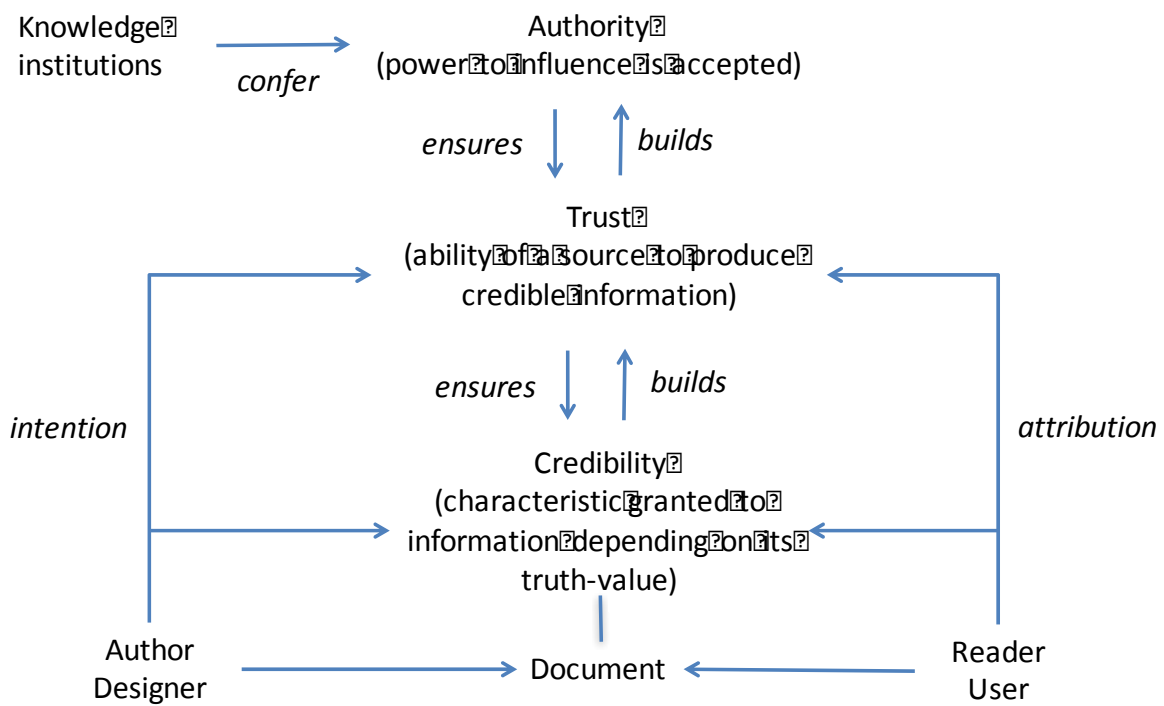

Figure 1: The Authority, Trust and Credibility (ATC) Model. With the downward process, knowledge institutions confer authority to a source, this authority ensures trust, which ensures the credibility of the information. With the upward process, the credibility of the information builds trust, which builds the authority of the source.

*Application of the model to Wikipedia*

The ATC model can help to understand the phenomena of authority, trust and credibility about Wikipedia. As stated above, Wikipedia is not initially a source linked to an institution of knowledge. So trust in this source is built by an upward authority process. Positive experience of its credibility in the majority of these users lead to a certain level of trust. However, as the various studies mentioned above have indicated, trust in the collaborative encyclopedia is not optimal. This contrasts with the high frequency of use of this source. An explanation can be found in the attitude of knowledge institutions to the collaborative encyclopedia. As seen above, the majority of teachers in high school and university have advised learners against the use of this source because, according to them this encyclopedia is untrustworthy. The encyclopedia has a rather bad academic reputation.

Going back to our model, we can say that there are effects at work in the opposite direction. The mostly positive experience of the source contrasts with its mostly negative reputation among those who represent the institution of education. The upward process of building up authority is somewhat prevented by its reputation in academia (fig. 2).

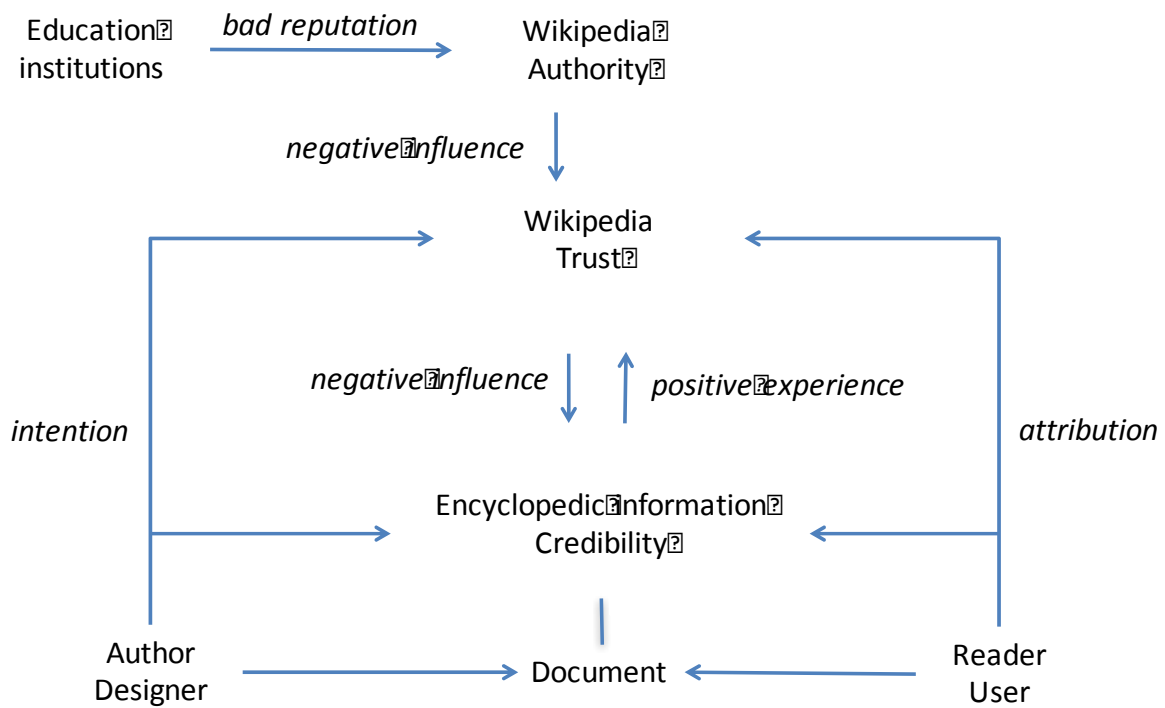

Figure 2. Wikipedia authority, trust and credibility. The educational institution can spread a bad reputation on Wikipedia, which decreases its authority, has a negative influence on its trust, which negatively influences the credibility of the information. Conversely, a positive experience of credibility of Wikipedia information increases readers' trust.

As a complement, we focus on citation of sources in Wikipedia (fig.3). The ATC model can help to understand why this practice became central to the community of wikipedians. By this way, the community tried to ensure the credibility of encyclopedic statements and to encourage a relation of trust with its readers. The editors show that they don't speak in their own names but that they are reporting an already published speech. Citing sources is a way of linking Wikipedia's statements to existing knowledge institutions. In relation to the ATC model, we can say that it is a matter of importing an external authority.

This process is thus meant to earn the reader's trust and thereby allow Wikipedia to gain authority. Yet as we saw earlier, trust in Wikipedia is not very high. One of the possible explanations lies with the implementation of the process. Many of the Wikipedia's articles do not have references. The contributors to the encyclopedia are aware of this fact since many « unreferenced » and « citation needed » annotations were added. In March 2016 we counted 87 173 "citation needed" tags in the French-speaking Wikipedia and 48 469 articles with an "unreferenced" template. Moreover, the sources that are cited are different in nature. Some of the sources are scientific, but other are from the state, the media, associations, and the social web (Luyt, Tan, 2010; Ford et al., 2013). Thus not all come from renowned institutions of knowledge. This discrepancy puts a limit on the effects of the process of importing authority that is at work in Wikipedia (fig. 3).

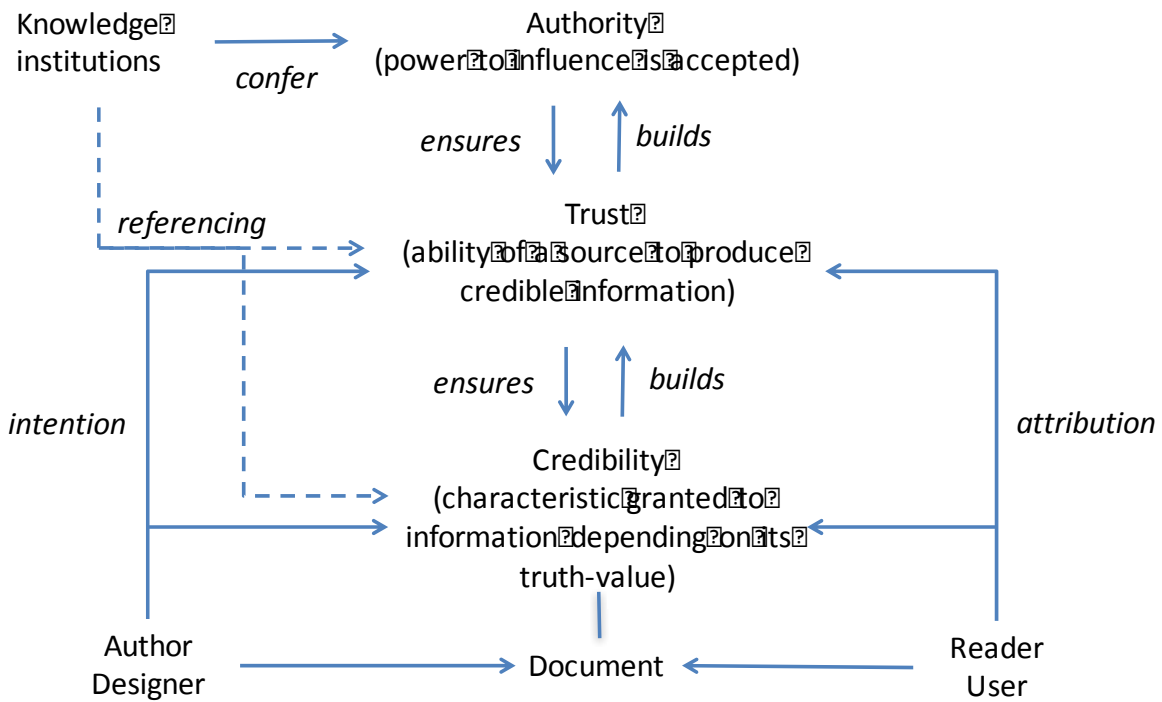

Figure 3. Referencing as importing authority to enhance trust and credibility. The use of references from knowledge institutions is a means to increase trust and credibility by importing the authority of knowledge institutions.

**Discussion**

Models are inherently reductive in that they fail to incorporate all the variables that can potentially influence the investigated process . However, it seems important to discuss whether our model is compatible with the results of past research on credibility and/or trust regarding online. We referenced several empirical studies on this topic and in particular on Wikipedia.

While the issue of authority was traditionally seen as a process intrinsically linked to knowledge institutions, the ATC model highlights the dynamic process of building the authority. Positive experiences with the source and its favorable reputation are two key factors that enable it to build trust. Several studies show that these two factors are important in the way people assess the credibility of information (Hilligoss, Rieh, 2008; Metzger, Flanagin, Medders, 2010). They are both at the origin of heuristics processes, *i.e.* pre-constructed rules available in its memory that minimize cognitive effort and time.

Using past experience with the source to trust it is widely spread heuristic (Rieh, 2002; Kelton, Fleischmann, Wallace, 2008; Hilligoss, Rieh, 2008). The positive experience of the source has led to test its expertise and honesty, thus dispensing with a thorough review of the credibility of the proposed information.

As several studies highlighted (Hilligoss, Rieh, 2008; Metzger, Flanagin, Medders, 2010), the reputation of a source is a social phenomenon that influences epistemic trust and is at the inception of heuristics of judgment in the context of online environment. The vehicles of reputation can be either informal or formal (Origgi, 2015). The former is made up of socio-cognitive phenomena linked to how information spreads such as rumors or recommendations: for instance, a teacher can recommend using a source to her or his students and thereby contribute to source's good reputation among this public. The latter (formal) are the information rating or ranking systems created by online algorithms. Online technologies indeed enable to aggregate the opinions of users and to translate them into an index. This can be found on trading websites, some forums, social network, and Facebook or Twitter (number of followers, tweets, re-tweets) (Westerman, Spence, Van Der Heide, 2012). These indexes are used to determine credibility, which is called tabulated credibility (Flanagin, Metzger, 2008) or aggregated trustworthiness (Jessen, Jørgensen, 2012). They are indicators of the upward process of building up authority.

Two remarks can be made on this subject. First, more detailed studies are needed to determine the extent to which the ATC model could be applied to different digital media such as blogs or social networks. Secondly, we also need some studies to evaluate the respective weight of the source and reputation experience. These two factors combine to participate in an upward dynamic of building up authority. But they can also be in tension as shown in the example of Wikipedia.

In order to complement the approach proposed by the ATC model, it seems necessary to add that the design of a site is an important factor of trust (Fogg et al, 2003; Robins & Holmes, 2008; Watson, 2014). The users often rely on visual aesthetics and on the appropriateness of design of the source (Choi, Stvilia, 2015). This seems compatible with our model if these elements are considered as possible characteristics of the source that create a feeling of trust towards it, which ultimately influences the credibility of the information.

The social and contextual dimensions of the phenomena linked with trust, authority and credibility should also be taken into account. The authority granted to a source fluctuates depending on social backgrounds or communities and the recognition of representatives of institutions of knowledge varies (Wilson, 1983; Mc Kenzie, 2003). Group-based credibility shows that opinions circulate in a group regarding whether sources of information can be trusted (Metzger, Flanagin, Medders, 2010). Moreover, the trust in a source depends on the topic at hand. Even for sources that deal with all types of subjects, trust can vary depending on whether the context of the research is recreational or scholarly (Sahut, 2014a). Thus much remains to be investigated to identify how the process of building authority works depending on social background and on how hierarchies of sources can be created depending on the nature of the information-seeking.

**Conclusion**

We initially noted that credibility, trust, and authority were three concepts frequently used to describe epistemic belief phenomena but also that it seemed difficult to define them and to determine the semantic relationships between them. This observation led us to propose both a model and definitions that link the three concepts.

This in turn led us to identify a downward process of acknowledging authority. Following that logic, the reader recognizes the characteristic of institutional authority embedded in the document. This leads the reader to trust the source as it vouches for the credibility of the information.

We also discerned an upward process of building up authority. Spreading information that is deemed credible contributes to establishing trust in a source even if it is not ostensibly part of an institution. The authority of a source is as great as the number of individuals who trust this source and the strength of this relationship. Authority is built in communities of different natures and sizes. The experience of the source plays a crucial role, as does its reputation. Shared opinions among a community over the reliability of a source, whether they are positive or negative, have an effect on its authority.

This model can represent the redistribution of cognitive authority. We used the example of Wikipedia which, despite a very large readership, is not (yet) recognized as an authority in the same capacity as printed encyclopedia. Among young people, the upward process of building up authority is hindered by a majority of negative opinions among teachers over the reliability of this source. The contributor community realized there was a reputation problem. They established rules around quoting sources in an attempt to fix it. In practice however, this process of importing authority is hindered by uneven referencing, both in quantity from article to article, and in the authority of the sources used as reference.

We would point out that the upward and downward processes work differently on a timescale. Downward authority guarantees instantaneous trust and credibility, obviously under the condition that the distinctive signs and known and recognized by the reader. Authority is build up with time. If it is deemed credible over time, a source is likely to capitalize on the trust it gets through its reputation. Thus it seems important to study the future evolutions of the trust and reputation of sources such as Wikipedia in order to better analyze the possible evolutions of how cognitive authority is set up.

**Acknowledgements**

The authors would like to thank Michael Buckland for his comments on earlier version of this paper.

**About the authors**

Gilles Sahut is teacher of documentation at the Toulouse School of Education (France) and researcher at the Laboratoire d'Études et de Recherches Appliquées en Sciences Sociales (University of Toulouse). He obtained his PhD in Information and communication sciences in 2015. His work is mainly focused the credibility and authority of Wikipedia and Media and Information Literacy.

André Tricot is professor of psychology at the Toulouse School of Education (France) and member of the Work and Cognition Lab. (French national council for research). In 2014-15, he was the head of the group that design grades 1, 2 and 3 new curricula for primary schools in France. He was also visiting Professor at the School of Education, University of New South Wales, Sydney. He obtained his PhD in Cognitive Psychology, Aix-Marseille University, France in 1995. André's work is mainly focused the study of human learning and information seeking in digital documents.